\documentstyle[12pt,epsfig,,prc,aps]{revtex}
\begin{document}

\draft
\title{Spinodal and dynamical instabilities at the phase 
transition
from the quark-gluon plasma to hadrons}

\author
{P.\
Bo\.{z}ek\footnote[1]{On leave from: Institute of 
Nuclear Physics, Cracow,
Poland}, Y.B.~He
and J. H\"ufner  }

\address{Institute for Theoretical Physics, University 
of Heidelberg,
Philosophenweg 19, D-69120 Heidelberg, Germany}

\date{\today}

\maketitle

\begin{abstract}
%\parbox{14cm}
The stability of an expanding  parton plasma is
analyzed within quasi-particle models.
The effective mass of the parton is calculated 
self-consistently
from a gap equation which
is either obtained from the Nambu Jona-Lasinio  
Lagrangian or from a fit to 
observables calculated on the lattice.
The latter shows an effective confinement. At thermal 
equilibrium the
stability is studied within thermodynamics (mechanical 
stability) and
via a 
linear response analysis of the Vlasov equation. The 
instabilities
related to a first-order phase transition are found. 
For a plasma expanding
in three and one dimensions far from equilibrium
a new type of instability, called dynamical, appears. 
The relation to
cluster formation is shown in a molecular dynamics 
calculation.
%show a dynamical instability
%for the gap equation obtained from the lattice data but 
%not for the Nambu
%Jona-Lasinio model. 
\end{abstract}
\pacs{PACS numbers: 25.75.-q, 12.38.Mh, 12.39.Ba}
\narrowtext

%\vfill
\newpage

\section{Introduction}

The experimental search for the quark-gluon plasma in 
ultra-relativistic
nuclear collisions requires the development of reliable 
and
predictive models of the formation of the plasma and of 
its hadronization
in order to be able to extract observables which signal 
the
existence of a parton plasma  in the high energy density 
region of the 
colliding nuclei. While lattice calculations have 
yielded important
information on the equilibrium properties of a parton 
plasma
(at least for baryon chemical potential $\mu=0$), they 
are unable yet
to describe the space-time development of the plasma 
formation and its
subsequent decay into hadrons.
The formation phase of the plasma is calculated usually 
in cascade
models \cite{cascade},
which predict high values for the
energy densities in the center of the collision.
These values should be sufficient, according to lattice 
QCD
calculations, for a plasma of deconfined quarks and 
gluons to be formed.
A satisfactory microscopic calculation of the 
hadronization process in
space-time is not yet available, in particular because 
of the complexity
of the confinement problem. In this situation model 
studies are
appropriate to elucidate the physics of the process. The 
present work is
one of them. 

In our work the partonic phase will be
described within  a quasi-particle model with an
effective mass.
The simplicity of the model
allows to describe at the same time the equilibrium 
properties and the
expansion dynamics. No hadronization is yet included. 
The medium dependent mass $m$ (in the case of thermal 
equilibrium $m(T,\mu)$,
and $m(x,t)$ for an expanding parton gas) is calculated 
self-consistently
from a gap
equation which relates the mass $m$ to the local parton 
distribution
function in phase space. The self-consistent mass itself 
determines the
momentum distribution in the case of thermodynamical 
equilibrium
or the evolution equation (the Vlasov equation) of the 
phase-space 
distribution in the nonequilibrium situation.
 We study two quasi-particle models,
which are in a sense complementary:
\begin{enumerate}
\item{} The gap equation is the one of the Nambu 
Jona-Lasinio (NJL)
model \cite{NJL,NJLrev} with non-zero current quark 
mass. For $\mu=0$
it shows a cross-over
transition around a ``critical'' temperature 
$T_c$  and a first one at low temperatures
and finite baryon density.
\item{} We introduce a phenomenological parton model 
(PPM):
Its gap equation is adjusted such that the 
quasi-particle model
reproduces the lattice data for
the energy density $\epsilon(T)$ and the chiral 
condensate $\langle 
{\bar \Psi} \Psi \rangle (T)$ around $T_c$ (for $\mu=0$)
\cite{inne}. It is described in Ref.~\cite{transp} and 
generalized to 
$\mu\neq 0$. 
It shows effective confinement in that the system
cannot expand indefinitely, because the effective
parton masses would grow to infinity.
\end{enumerate}
For these two
quasi-particle models we study  the stability of the 
system in 
three physical situations.
\begin{itemize}
\item{}Within equilibrium thermodynamics  the 
region in the plane of temperature $T$ and 
baryon density $\rho_B$ is
calculated
where $\partial P/\partial \rho_B <0$,
{\it i.e.}, the system is mechanically
unstable.
At the same time  the region of the mixed phase in the
$T$-$\rho_B$ plane is calculated which indicates
 the presence of a first order phase transition.
\item{} Using the method of linear response, the 
stability of
the space-time dependent Vlasov equation around thermal 
equilibrium
is investigated and 
the exponential growth rates are determined for the 
region of 
instability.
\item{} The stability analysis of the Vlasov equation is 
also applied to
situations far from equilibrium as they occur in 
three-dimensional (Hubble) and one-dimensional (Bjorken) 
expansions
 of a system of partons. 
%In both cases instabilities are observed at 
%$\rho_B=0$ for
%the phenomenological parton model but not for the NJL model.
\item{} In order to show that the instabilities are 
related to hadronization,
we have studied the expansion and clusterization 
of an initially compact system of partons via
molecular dynamics.
% using effective medium dependent masses.
%After sufficient time the fireball fragments into smaller clusters 
%containing at least two partons.
\end{itemize}

\section{The model}

\label{model_sec}

\begin{figure}[b] 
\begin{center}
\epsfxsize=6in
\epsffile{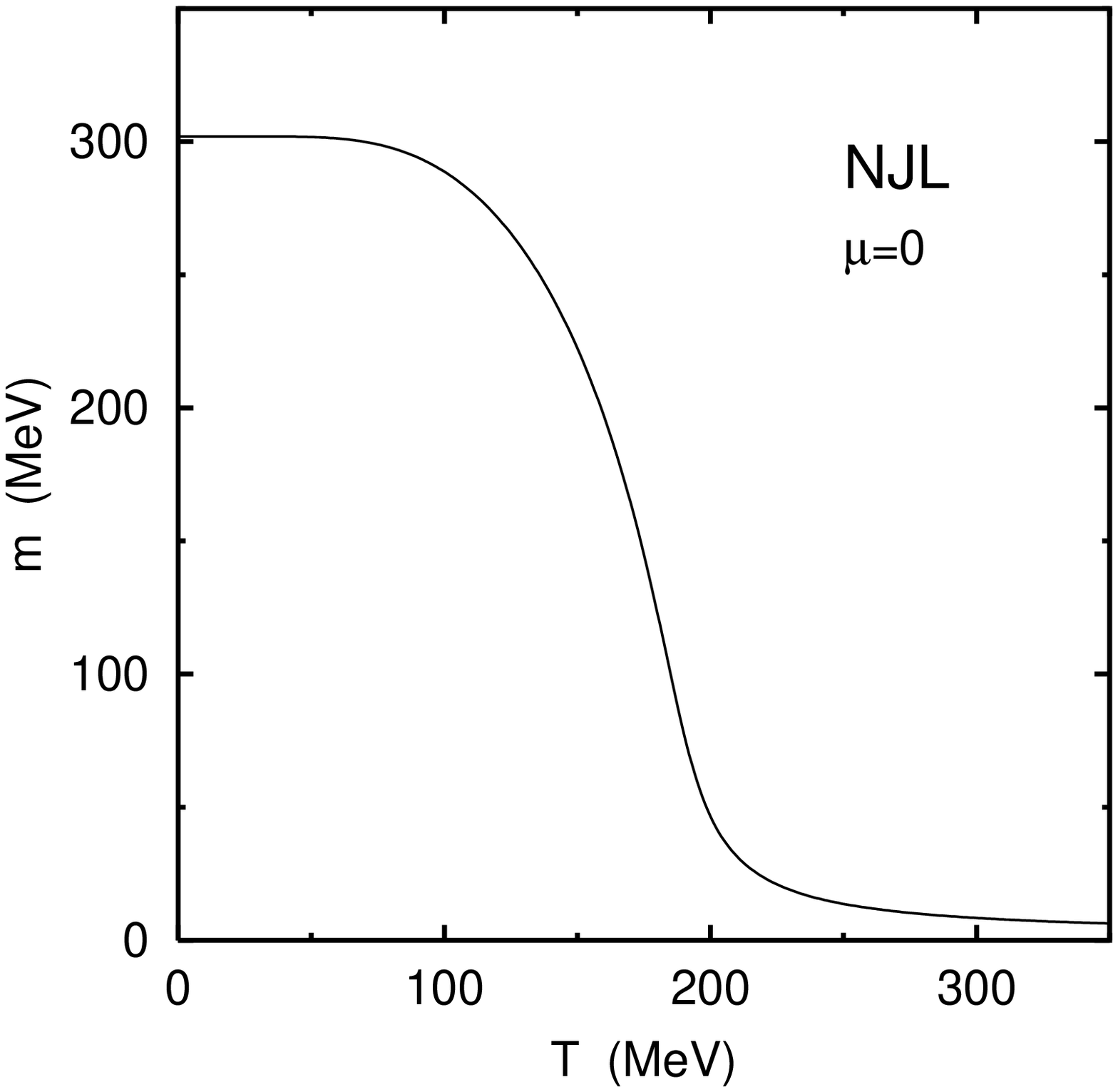}
\caption[]{The temperature dependence of the parton mass
in the two-flavor NJL model for $\mu=0$.}
\label{mtNJL_fig}
\end{center} 
\end{figure}

\begin{figure}[b] 
\begin{center}
\epsfxsize=6in
\epsffile{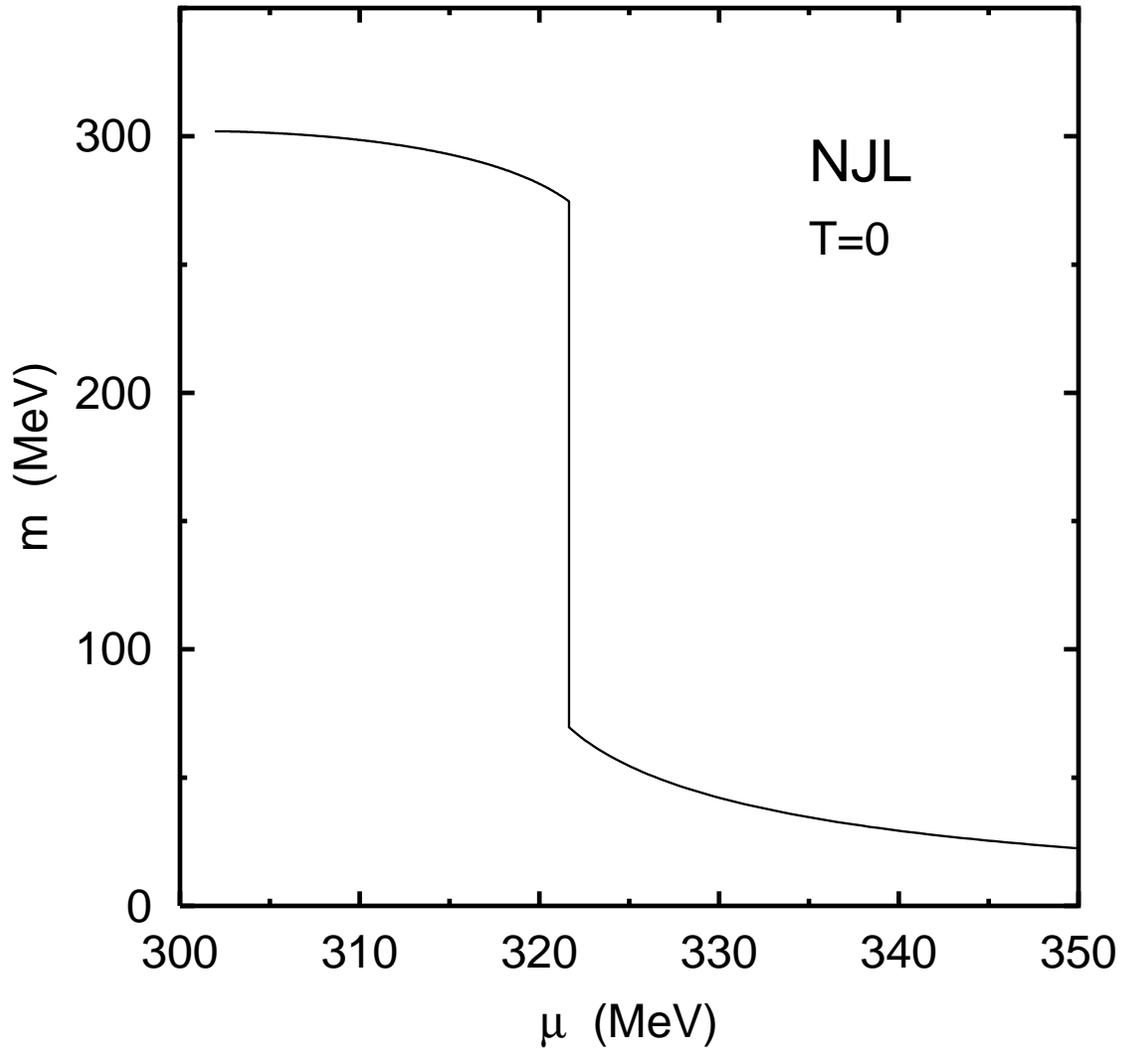}
\caption[]{The dependence of the parton mass on the 
chemical potential 
in the NJL model at $T=0$. Note the expanded scale on 
the abscissa.}
\label{mfdNJL_fig}
\end{center} 
\end{figure}

The basic assumption of our work is the quasi-particle 
approximation for
a gas of ``partons'' (quarks and gluons) with an 
effective medium dependent
mass \cite{inne,transp}.
The thermodynamic potential of the 
quasi-particle system composed of fermions, antifermions
and gluons with degeneracies $g_f$, $g_{\tilde{f}}=g_f$ 
and $g_g$,
respectively, is
\begin{equation}
\label{thpot}
P= \int\frac{d^3p}{(2 \pi)^3} \frac{p^2}{3E} \Bigg[ g_f 
f_f(p)+
g_f f_{\tilde{f}}(p) + g_g f_g(p) \Bigg] - V(m) \ ,
\end{equation}
where $f_i$ is a Fermi or Bose distribution function and 
$V(m)$ is the mass
dependent bag pressure.
In our calculation we use the {\em same} effective mass 
for quarks and gluons.
The parton mass depends on the medium through the gap 
equation~:
\begin{equation}
\label{gap}
\frac{dV}{dm}=-\int\frac{d^3p}{(2 
\pi)^3}\frac{m}{E}\Bigg[ g_f f_f(p)+
g_f f_{\tilde{f}}(p) + g_g f_g(p) \Bigg]  \ ,
\end{equation}
which follows from the thermodynamical relation
\begin{equation}
\label{simplegap}
\frac{\partial P}{\partial m} = 0 \ .
\end{equation}
Eq.~(\ref{thpot}) is not the most general expression for 
the 
pressure. For instance the bag
pressure could also depend on the baryon density. 
Since lattice data are only available for zero baryon 
density, we cannot 
explore this possibility.
For the NJL model, which has no gluons,
 the explicit form for $dV/dm$ can be derived from
the NJL Lagrangian within the mean field approximation.
One has 
\begin{equation}
\frac{dV}{dm} \equiv \frac{(m-m_0)}{2G} -  g_f
\int_{|p|<\Lambda} \frac{d^3p}{(2\pi)^3} \frac{m}{E} \ ,
\end{equation}
where  $G$ is the four fermion coupling constant, 
$\Lambda$ a
three-momentum
cutoff, and $m_0$ the current quark mass. At zero baryon
density the NJL model predicts a second order phase 
transition (for $m_0=0$)
or a crossover behavior (for $m_0>0$) (see 
Fig.~\ref{mtNJL_fig}).
At finite baryon density and low temperature the systems 
shows a first
order phase transition for moderate values of the 
current quark mass
(Fig.~\ref{mfdNJL_fig}).  We use the parameters 
$m_0=2$~MeV,
$\Lambda =653$~MeV and $G\Lambda^2 =2.1$.

\begin{figure}
\begin{center}
\epsfxsize=6in
\epsffile{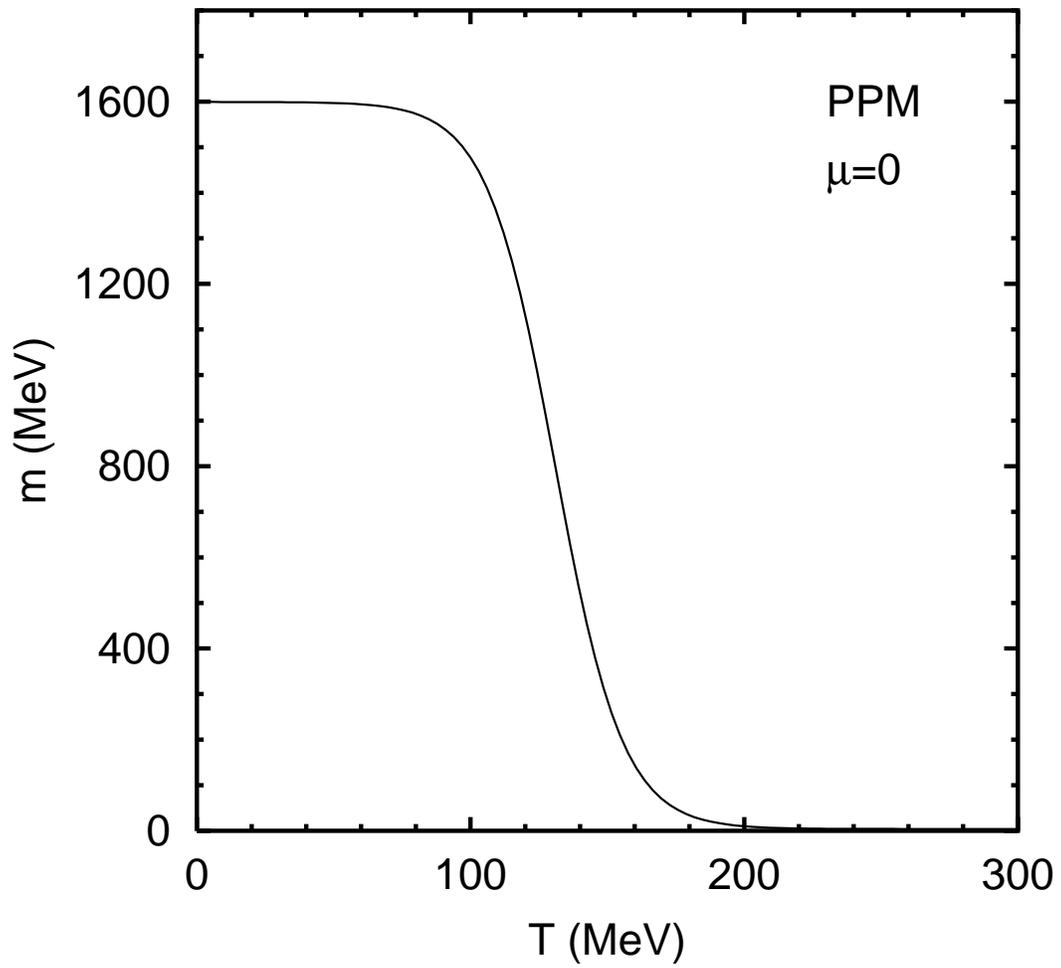}
\caption[]{The temperature dependence of the parton mass 
in the phenomenological parton  model (PPM) at  
$\mu=0$.}
\label{mt_fig}
\end{center} 
\end{figure}

\begin{figure}
\begin{center}
\epsfxsize=6in
\epsffile{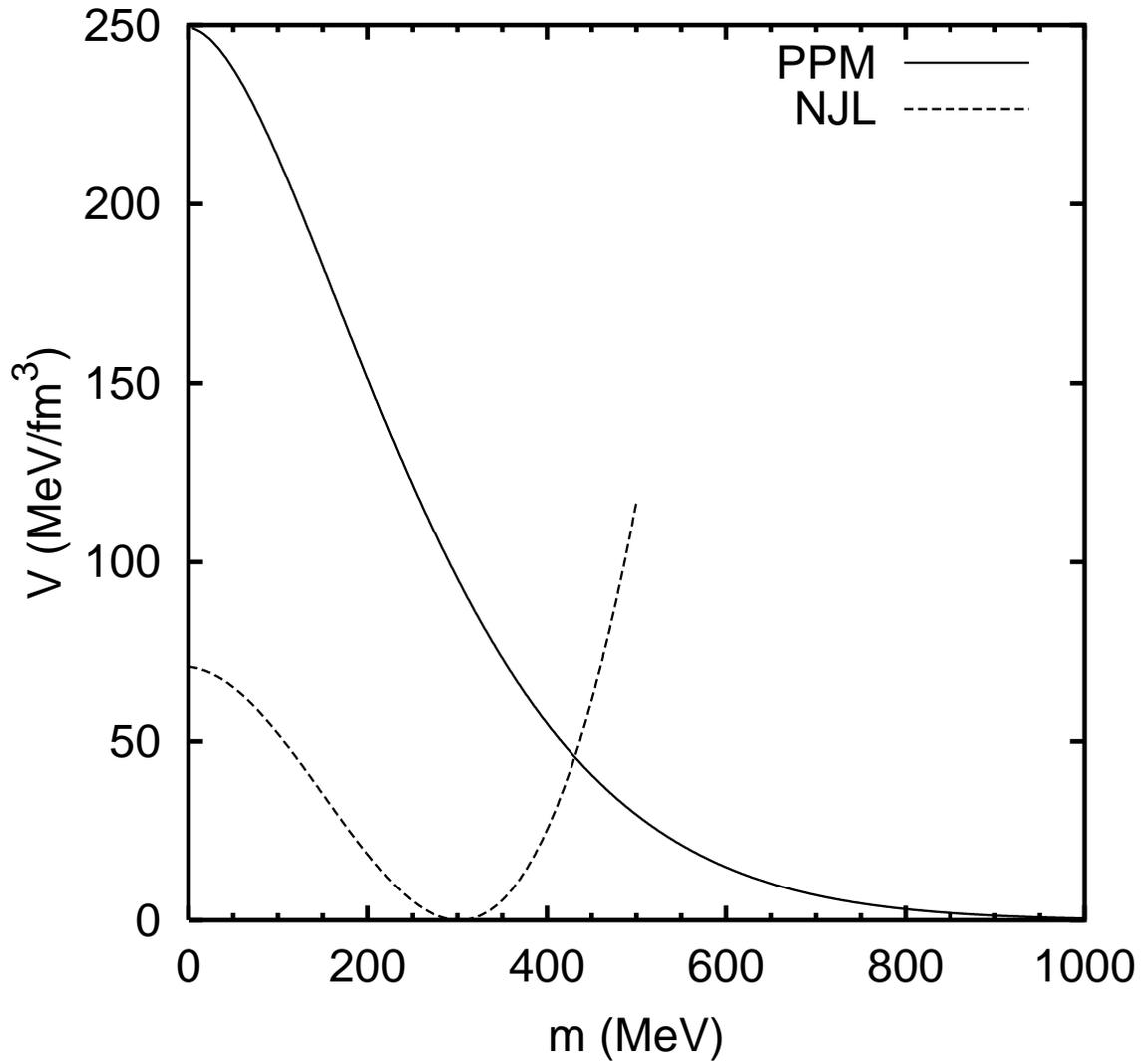}
\caption[]{The bag pressure density $V$ as a function of 
the effective mass 
%determines the porperties of the quasi-particle model. 
for the NJL model (dashed curve), 
and for the phenomenological parton model (solid 
curve).}
\label{vofm_fig}
\end{center} 
\end{figure}

\begin{figure} 
\begin{center}
\epsfxsize=6in
\epsffile{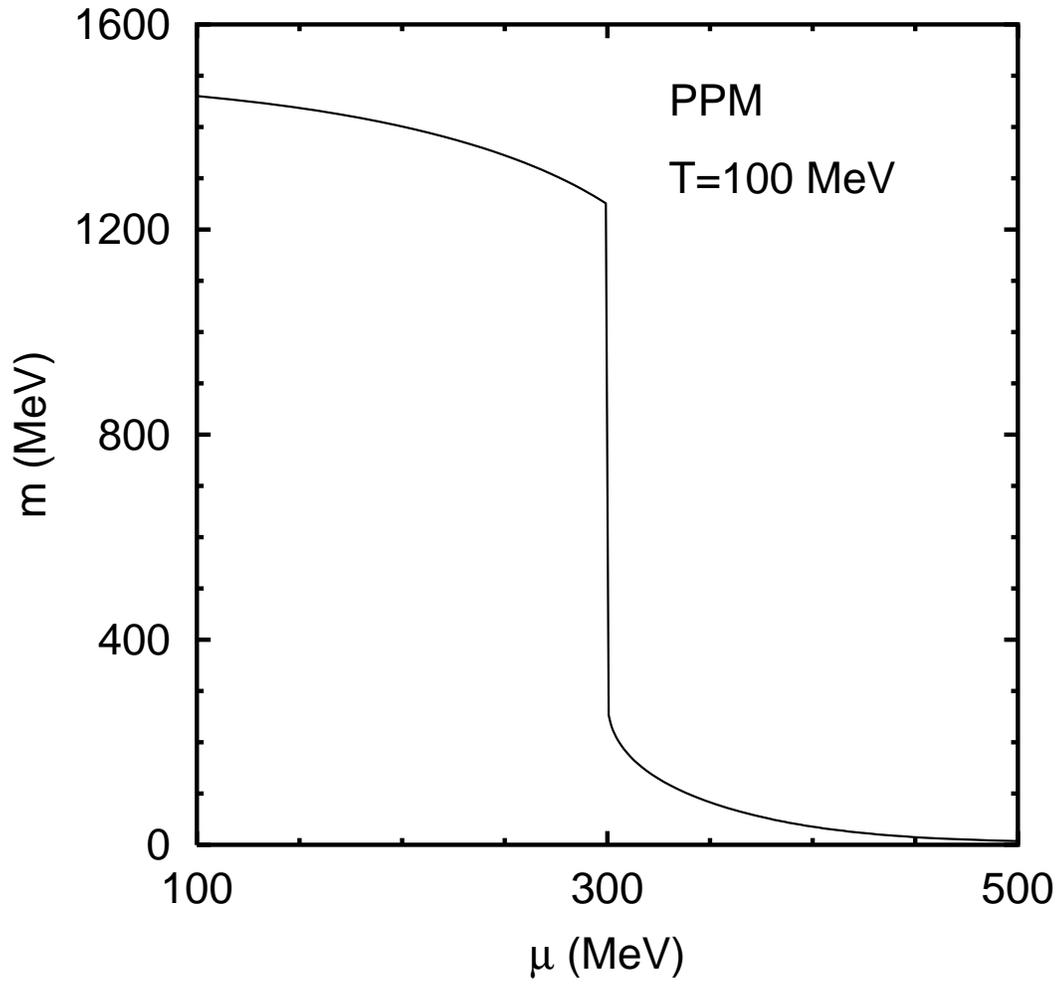}
\caption[]{The dependence of the parton mass on the
chemical potential 
in the 
phenomenological parton  model
at $T=100$~MeV.}
\label{fd_fig}
\end{center} 
\end{figure}

For the second model, called the phenomenological parton 
model 
(PPM)\cite{transp}, the temperature dependence of the 
parton mass 
$m(T)$ is obtained by fitting the energy density
\begin{equation}
\epsilon= \int\frac{d^3p}{(2 \pi)^3} E \Bigg[ g_f 
f_f(p)+
g_f f_{\tilde{f}}(p) + g_g f_g(p) \Bigg] + V(m) 
\end{equation}
to lattice QCD data around  $T_c$.
The parton mass in the 
high temperature region $T\gg T_c$ is
required to go smoothly towards zero in order to 
reproduce qualitatively
 the vanishing chiral condensate at high temperature.
%Th requirement to reproduce the temperature dependence of the parton
%mass $m(T)$ defines the gap equation for this model.
For the present study we use the same procedure to 
obtain
$m(T)$
as explained in Ref.~\cite{transp},
% (Fig.~\ref{mt_fig}), 
except that here 
we use Fermi and Bose
distributions for the fermions and gluons, respectively, 
and take
$g_f=12$ corresponding to the two-flavor QCD. 
The energy scale is  set by the critical temperature 
$T_c=140$~MeV for the
two-flavor QCD.
Fig.~\ref{mt_fig} shows $m(T)$.
Once it is known, one can use Eq.~(\ref{gap}) to obtain 
$dV/dm$
and $V$ as a function of $m$.
As explained in the beginning of this section,
the quasi-particle model is completely defined once the 
bag pressure $V(m)$ is given.
Fig.~\ref{vofm_fig} compares the bag pressure $V(m)$ of 
the 
PPM model with the one of the NJL model. Note that we
use 
$g_f=g_{\bar f}=12$ in both models for the fermions, but
for the gluons $g_g=0$ in the NJL and $g_g=16$ in the 
PPM model.
%Once $dV/dm$ is determined at $\mu=0$,
The use of the Fermi and Bose distribution functions and 
the assumption
that $V$ depends only on  $m$ allow  one to 
explore the parton mass and the thermodynamical 
properties of the system
 in the whole plane of temperature and baryon density, 
even
where there are no lattice data. 
Fig.~\ref{fd_fig} shows the parton mass as a function of 
the
chemical potential $\mu$ at $T=100$~MeV for the 
phenomenological
parton model. A first order phase
 transition is present at finite baryon density although 
at
 zero baryon density
we have, following the lattice QCD, only a crossover
 behavior around $T_c$. This behavior, crossover 
transition at $\mu=0$
and first order one at finite $\mu$, is also found for 
the NJL model.

The quasi-particle models, whose thermodynamical 
properties are described
above, can also be 
used to describe the space-time evolution of parton 
system. 
The underlying equation is the Vlasov one
\begin{equation}
\label{vlasov_equ}
\partial_t f_i(t,x,p)+\frac{\vec p}{E(t,x,p)} \cdot 
\nabla_x f_i(t,x,p)
- \frac{m(t,x)}{E(t,x,p)} 
\nabla_x m(t,x)\cdot  \nabla_p f_i(t,x,p)=0 \ ,
\end{equation}
where the quark, anti-quark or gluon  distribution 
functions
$f_i$ at time $t$, at space point $x$ and
 at three momentum $p$ describe on-shell partons with 
energy
 $E=\sqrt{m^2+{\vec{p}}^2}$. 
In a quasi-particle model there are no interactions 
between the 
quasi-particles, and therefore there are no collision 
terms on the 
r.h.s. of Eq.~(\ref{vlasov_equ}).
The space-time dependent mass $m(t,x)$ is calculated
 self-consistently from the gap equation
  as a function of the non-equilibrium 
parton distribution functions $f_i(t,x,p)$~:
\begin{equation}
\label{gapgen_equ}
 \frac{dV}{dm} = - \sum_i g_i  \int \frac{d^3p}{(2 
\pi)^3}
\frac{m(t,x)}{E(p,t,x)} f_i(t,x,p)  \ .
\end{equation} 
Eqs.~(\ref{vlasov_equ}) and (\ref{gapgen_equ}) have to 
be
solved simutaneously
for an evolving parton gas. This has been done already 
for the 
phenomenological parton model with the result that
during the expansion the partons acquire high
masses and the 
expansion stops \cite{transp}. We call this behavior 
``effective 
confinement''.

\section{Mechanical stability and first order phase 
transition}

For a system at thermodynamic equilibrium 
the mechanical instability is defined 
by the condition
\begin{equation}
\label{ms}
\Bigg(\frac{\partial P}{\partial \rho_B}\Bigg)_T<0 \ .
\end{equation}
With the gap equation~(\ref{simplegap})
the condition (\ref{ms}) 
is equivalent to
\begin{equation}
\left( \frac{\partial \mu}{\partial \rho_B} \right)_T < 
0 \ .
\end{equation}

The negative derivative of the pressure or the chemical 
potential with respect to the density
signals an instability of the homogeneous state of the 
system.
Generally such an instability is related to a first 
order phase
transition \cite{ll}. 

\begin{figure}[b] 
\begin{center}
\epsfxsize=6in
\epsffile{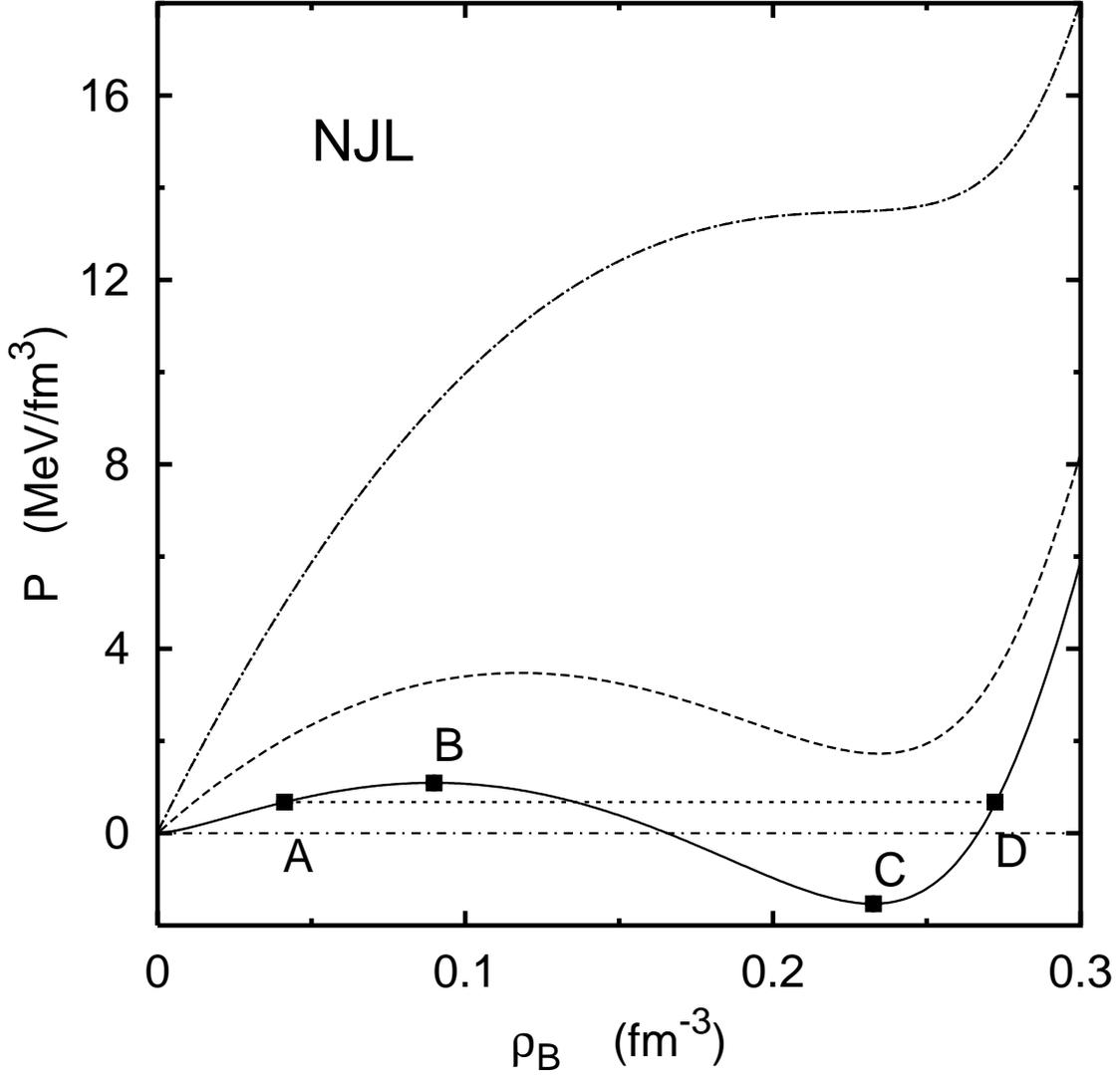}
\caption[]{The pressure density as a function of the 
baryon density in
the NJL model for 
$T=0$, $20$ and $45$~MeV, solid, dashed and 
dash-dotted lines, respectively.
The parts A--B and C--D of the $T=0$  curve
correspond to the  metastable states and the part B--C
to the unstable state of the system. The straight line 
from A to D is the 
Maxwell construction for $T=0$. A similar classification 
into unstable
and metastable regions can be made for $T=20$~MeV. For 
$T=45$~MeV the system
shows no first order transition.} 
\label{pressNJL_fig}
\end{center} 
\end{figure}

\begin{figure}[b] 
\begin{center}
\epsfxsize=6in
\epsffile{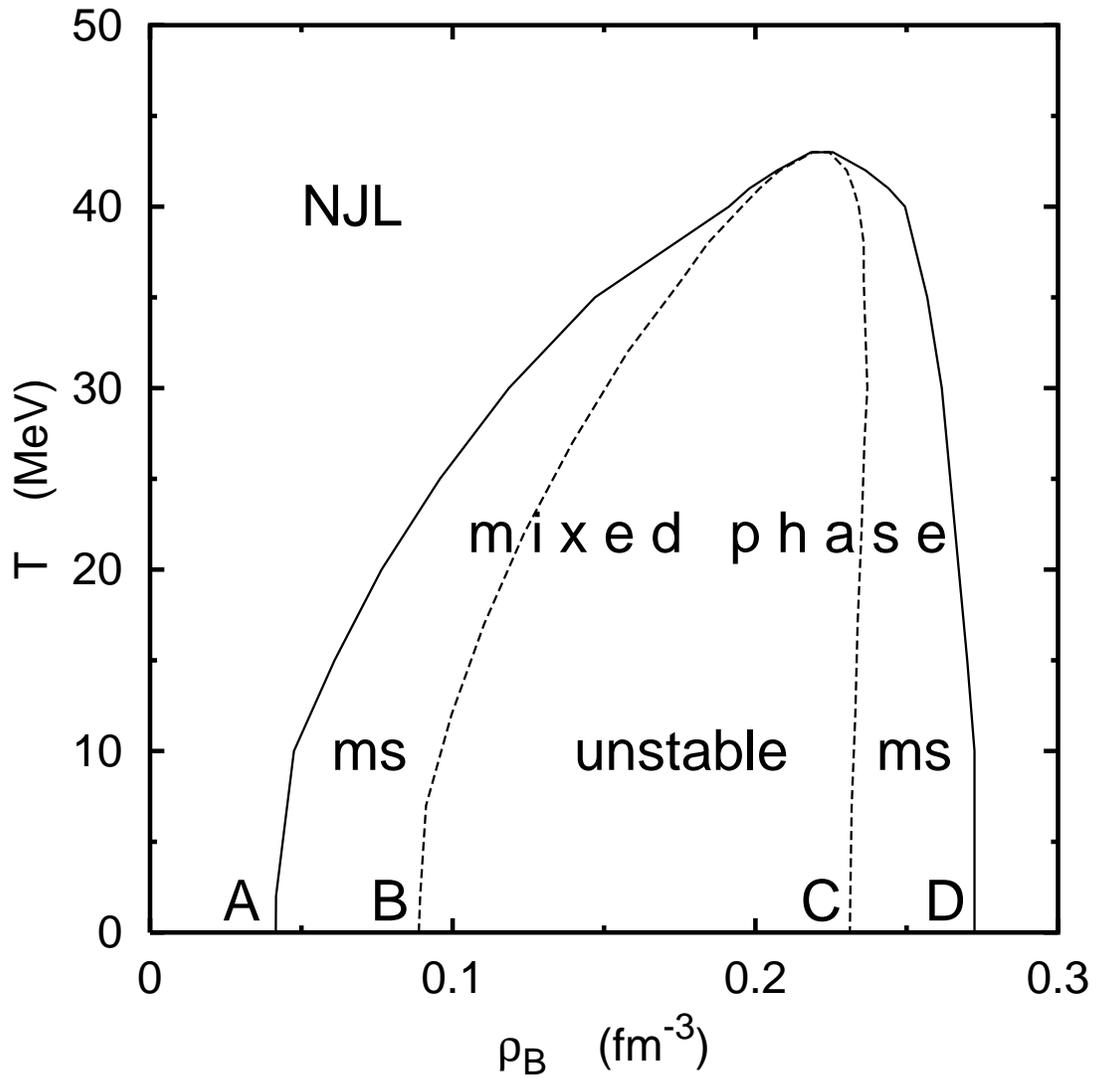}
\caption[]{The solid line in the $T-\rho_B$ plane 
separates the 
stable region
and the mixed phase. The mixed phase is devided into  
regions of instability (``unstable'') and metastability 
(``ms'').
The letters A to D at the bottom relate to $T=0$ and 
correspond to 
the points in Fig.~\ref{pressNJL_fig}. The results are 
for the NJL model.}
\label{NJLins_fig}
\end{center} 
\end{figure}

We discuss the numerical results for the two models, and 
begin with the NJL model. The pressure as a function of 
the baryon 
density is shown in Fig.~\ref{pressNJL_fig}
 for several values of the temperature.
For $T=0$ for example 
the pressure is not a monotonic
function of the density and a first order phase 
transition occurs.
For the interval B--C in Fig.~\ref{pressNJL_fig} the 
system is unstable
since one has 
 $\partial P/\partial \rho_B<0$.
The thermodynamical system is
in a metastable state in the  regions (intervals A--B 
and C--D in
Fig.~\ref{pressNJL_fig}) around the unstable region. 
Performing a Maxwell construction amounts to choosing 
the
correct solution of the gap equation for certain values 
of $\mu$ where
the gap equation has 2 or 3 solutions for the mass 
(short dashed line
in Fig.~\ref{pressNJL_fig}). The so constructed mass as 
a
function of the chemical potential has  a discontinuity 
at the critical
chemical potential $\mu_c$ (see Fig.~\ref{mfdNJL_fig}). 
The energy density and the
baryon density are also discontinuous at $\mu_c$. The 
discontinuity in 
$\rho_B$ means that the whole region AD of the baryon 
densities is excluded
from the phase diagram. This region is called mixed 
phase and is shown in
Fig.~\ref{NJLins_fig} in the plane of temperature and 
baryon density.
The region of mechanical instability defined by 
Eq.~(\ref{ms}) is a 
part of the mixed phase.
Inside the region of mechanical instability the system 
is
unstable with respect to non-homogeneous perturbations. 
In the whole
region of mixed phase it is preferable for the system to 
separate into
zones belonging to the  low or high density 
thermodynamically stable
phases.

\begin{figure}[b] 
\begin{center}
\epsfxsize=6in
\epsffile{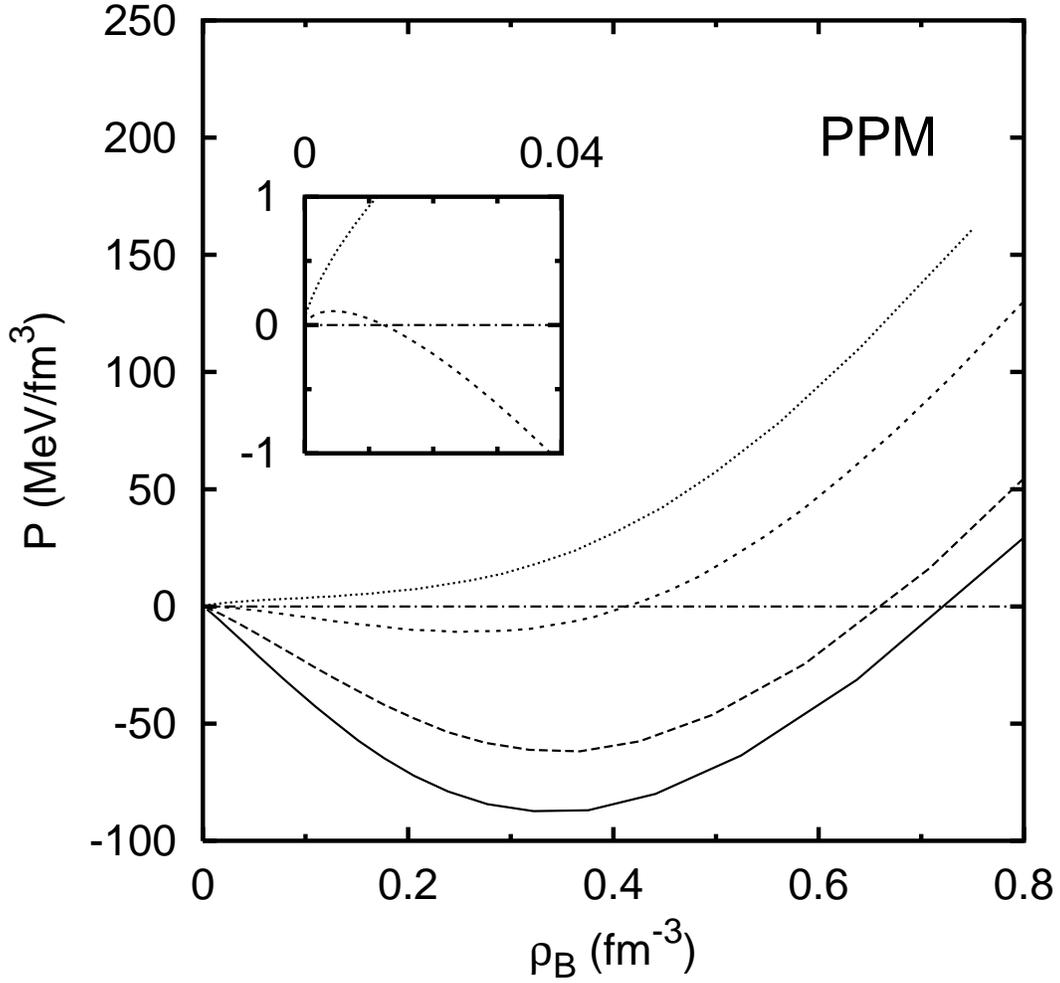}
\caption[]{The pressure density as a function of the 
baryon density in
the phenomenological parton model at different 
temperatures. 
The four curves (from bottom to top) correspond to 
temperatures 
$T= 0$, 50, 100, 120~MeV, respectively,
while the insert magnifies 
the details around the origin for temperatures $T=100$ 
and 120~MeV.}
\label{press_fig}
\end{center} 
\end{figure}

For the phenomenological parton model,
Fig.~\ref{press_fig} shows the pressure as a function of 
the baryon
density for several temperatures.
The behavior is qualitatively similar to the NJL model: 
For $T<120$~MeV
the pressure is not a monotonic function of the
density, and  
a Maxwell
construction is required. 
However, quantitatively the situation for the PPM model 
is rather 
different and we have chosen the insert to show the 
situation for small
$\rho_B$.

\begin{figure}[b] 
\begin{center}
\epsfxsize=6in
\epsffile{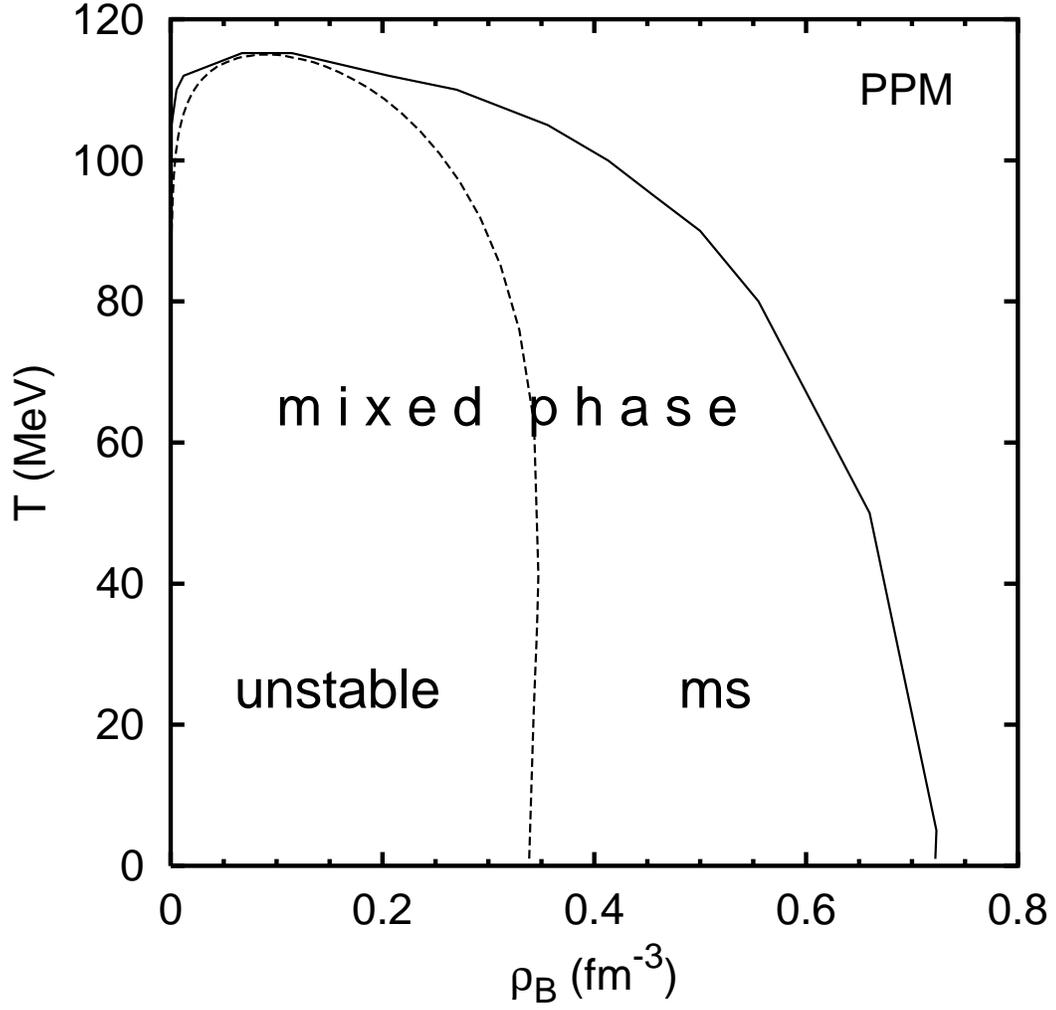}
\caption[]{The regions of the mixed, unstable and 
metastable phases in the
$T-\rho_B$ plane 
in the phenomenological parton model. Note that the 
whole line 
$\rho_B=0$ is part of the stable region.}
\label{phase_fig}
\end{center} 
\end{figure}

The regions of mechanical instability and metastability 
in the $T$-$\rho_B$ plane
are shown in Fig.~\ref{phase_fig} for the 
phenomenological parton model.
A region with temperatures  below $120$~MeV and
finite baryon density is unstable. As expected, for 
large temperature and/or 
baryon density the system is stable against mechanical 
perturbation,
which is natural since then the partons are almost 
massless.
However, the region around the $y$-axis ($\rho_B=0$ and 
$T\geq 0$)
is also stable.

The phenomenological parton  model 
 becomes questionable at values of $T$ and $\rho_B$ 
where  the parton
 masses become much larger than the hadron masses, since 
hadronization is not accounted for. This is the case for 
the interior of the 
region of the mixed phase.
 However, close to the boundary of this region the model 
should correctly describe the thermodynamic situation 
including the effect
of confinement.

Let us imagine a heavy ion collision and suppose that 
the system evolves 
at or close to thermal equilibrium:
A region of deconfined parton plasma is created 
at high temperature and high baryon density in the 
stable regions of the 
phase diagrams in Figs.~\ref{NJLins_fig} or 
\ref{phase_fig}.
During its expansion,
the system reaches the region in the $T$-$\rho_B$ plane 
where the local homogeneous parton distribution is 
unstable. 
In the spinodally unstable region
any perturbation of the homogeneous distribution 
amplifies
 and the system breaks up into regions (``clusters'') 
with
larger density. 
This is the onset of hadronization.
However, our quasi-particle models are not realistic in 
these regions,
their 
and the structure, sizes
and masses of the 
clusters cannot be reliably calculated.
%predicted in a simple local mean field model. 
%Moreover usually such clusters would be excited and should decay
%successively into hadrons. (
Sect.~\ref{md_sec} presents an illustration of the 
clusterization
process.

\section{Linear instability of the Vlasov equation near 
thermal equilibrium}

\label{linstab}

 In the semi-classical limit the collisionless expansion 
of the  
 plasma can be described by the  Vlasov equations 
(\ref{vlasov_equ})
together with the gap equation (\ref{gapgen_equ}).
In this section we address the linear stability problem 
of the Vlasov 
equation for the quasi-particle parton model around the 
 equilibrium solution.
We consider a small perturbation $\delta f_i$ around the 
homogeneous 
solution $f_i$ of the Vlasov equation~:
\begin{equation}
f_i(t,x,p)=f_i(p)+\delta f_i(t,x,p) \ ,
\end{equation}
where the index $i$ denotes the fermion, anti-fermion or 
gluon distribution.
%The homogeneous distributions $f_i(p)$ are solutions of the
%Vlasov equations 
%and by the gap equation they define the unperturbed
%space-time independent mass $m$.
%and are Fermi or Bose distributions with effective mass $m(T,\mu_B)$.
As it is standard practice \cite{nuclear,peth}, one 
linearizes 
the Vlasov equation and takes the Fourier transform 
in the  $t$ and $x$ variables  to obtain
\begin{equation}
\label{linvl_equ}
({\vec k}\cdot {\vec v} -\omega)\delta f_i(k,\omega,p) - 
\frac{m}{E}\delta m(k,\omega)
{\vec k}\cdot {{ \nabla}_p}f_i(p)=0 \ ,
\end{equation}
where  ${\vec v}={\vec p}/E$, and 
$\delta m(k,\omega)$ is the change of the mass induced 
by 
the perturbation of the distribution functions $\delta 
f_i$, and 
has to satisfy the 
 linearized gap equation
\begin{equation}
\delta \frac{dV}{dm}(k,\omega)=-\delta
\bigg[ \sum_i g_i \int \frac{d^3p}{(2 \pi)^3}
\frac{m}{E} f_i(k,\omega,p) \bigg]\ .
\end{equation}
This leads to
\begin{equation}
\frac{d^2V}{dm^2}\delta m(k,\omega)
= -\sum_i g_i \Bigg[ \int \frac{d^3p}{(2 \pi)^3} 
\frac{p^2}{E^3}
f_i(p) \ \delta m(k,\omega) + \int \frac{d^3p}{(2 
\pi)^3}
\frac{m}{E} \delta f_i(k,\omega,p) \Bigg] \ .
\end{equation}
The  fluctuation of the phase-space distributions 
$\delta f_i(\omega,k,p) $
can be expressed by the mass fluctuation $\delta 
m(k,\omega)$ using the
linearized Vlasov equation (\ref{linvl_equ})
\begin{equation}
 \delta f_i(\omega,k,p) = \frac{m}{E}\frac{\partial 
f_i(E) }{\partial E}
\frac{{\vec k}\cdot {\vec v}}{{\vec k}\cdot {\vec 
v}-\omega} \delta m(k,\omega)\ 
\end{equation}
and is inserted  into the linearized gap equation:
%One
%obtains a dispersion relation between the frequency $\omega$ 
%and the wave-vector $k=|{\vec k}|$ of the perturbation $\delta m(k,\omega)$
\begin{equation}
\Biggl[ \frac{d^2V}{dm^2}
+ \sum_i g_i \biggl[ \int \frac{d^3p}{(2 \pi)^3} 
\frac{p^2}{E^3}
f_i(p)  +  \int \frac{d^3p}{(2 \pi)^3}
\frac{m^2}{E^2} \frac{\partial f_i(E) }{\partial E}
\frac{{\vec k}\cdot {\vec v}}{{\vec k}\cdot {\vec v}
-\omega}\biggr] \Biggr] \delta m(k,\omega)=0 \ .
\end{equation}
The  propagation of the disturbance $\delta m$ is only
possible, $\delta m\neq 0$, 
if $\omega $ and $k$ are related, $\omega (k)$, such 
that the term in the square
brackets is zero~:
\begin{equation}
\label{disp_eq}
\frac{d^2V}{dm^2}
+ \sum_i \frac{g_i}{2 \pi^2}\biggl[ \int_0^{\infty} 
\frac{p^4dp}{E^3}
f_i(p)  +\int_0^{\infty} \frac{m^2p^2dp}{E^2}
 \frac{\partial f_i(E) }{\partial E}
\chi\bigg(\frac{\omega}{kv}\bigg)\bigg]=0 
 \ ,
\end{equation}
where 
\begin{equation}
\chi(s)=1-\frac{s}{2}\ln\bigg(\frac{s+1}{s-1}\bigg) \ .
\end{equation}
The dispersion relation $\omega (k)$ defined by
Eq.~(\ref{disp_eq}) has two general properties.
The  solutions for the frequency $\omega$ come in pairs 
$\pm\omega$. The frequency is proportional to the 
wave-vector
$\omega_k\sim k$, which is a general property as long as 
there is no scale
in the dynamical equations. In our models the absence of 
scale is due to the 
fact that the relation between the
density and the mass is local (zero-range interaction).

In the case of an imaginary solution of 
Eq.~(\ref{disp_eq})
for the frequency,
$\omega_k= i \Gamma_k$,  the initial
disturbance grows exponentially in time 
\begin{equation}
\delta f(x,t,p) \sim exp(\Gamma_k t) \ , 
\end{equation} 
i.e. an instability appears.  As
a consequence the system breaks up into smaller pieces.
The fact that the instability rate is proportional to 
the wave-vector 
($\Gamma_k \propto k$) means that the instability rates 
are unbounded,
becoming arbitrary large for large $k$ or small sizes.  
On the other hand we
expect that the true clusters into which a low density 
parton system
breaks up are the hadrons. The preference of the nature
for the two or
 three quark clusters cannot be described in our simple 
model
without color and with only local mean field 
interaction.
A finite range
 of  interaction or a minimal size of the cluster would 
 regularize this unphysical behavior. In that case
 there exists a maximal growth rate of instabilities for 
a  specific
 wave-vector $k$ \cite{nuclear}.
We note that the presence of  higher gradients in the 
dynamical
equations
would
 change the dispersion relation
 and limit the growth of excitations with large 
$k$\cite{grad}.
In our case, the wave-vector for the physically most 
interesting
instability, namely hadron formation,
 should be of the order of $1.0$~fm$^{-1}$. 
%For larger $k$ the perturbation probes scales
%on which the confinement is no longer effective and on which the hadron
%structure should appear.

\begin{figure}[b] 
\begin{center}
\epsfxsize=6in
\epsffile{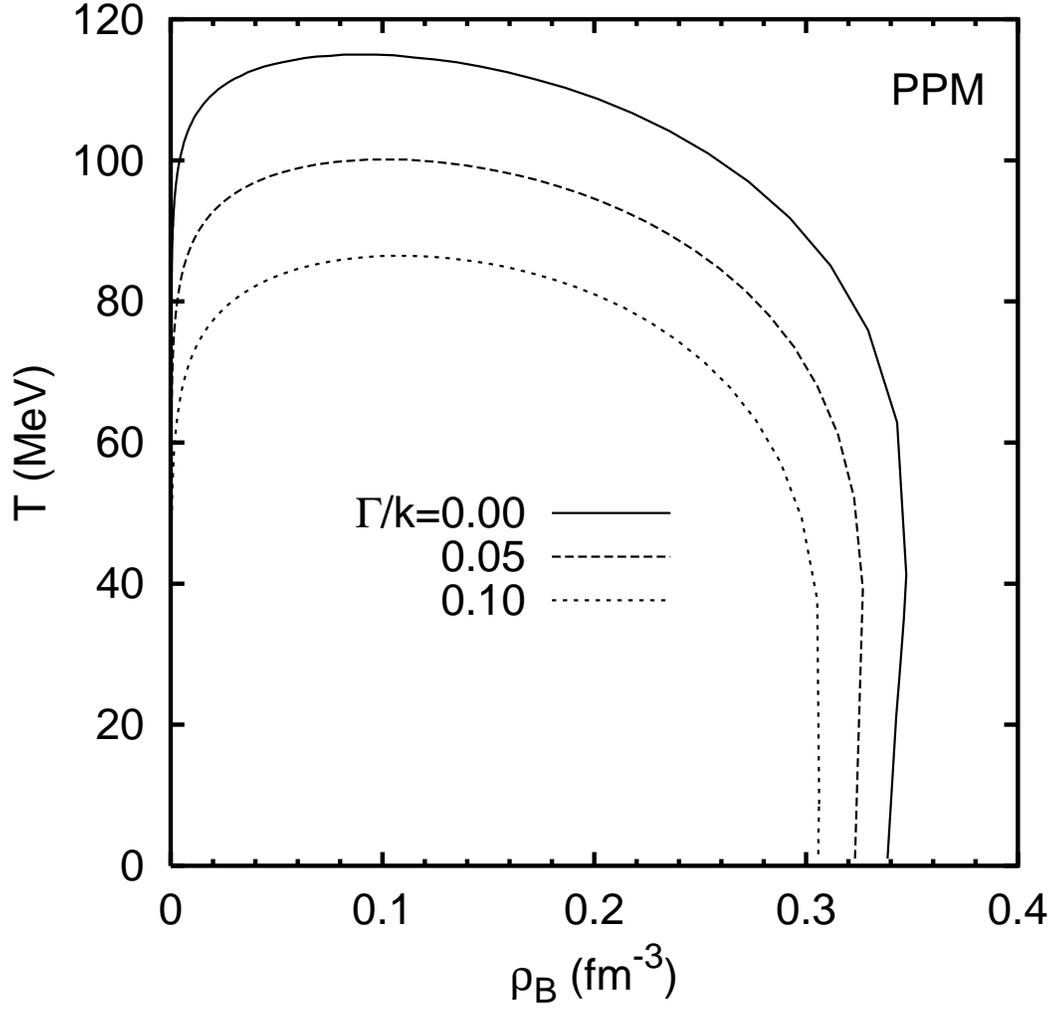}
\caption[]{The contour plots for the instabillity growth 
rate $\Gamma$ 
divided by the wave vector $k$
in the $T$-$\rho_B$ plane for the phenomenological 
parton model. The solid line
of this figure corresponds to the dashed one in the 
previous figure, which is 
the border of mechanical instability region.}
\label{rate_fig}
\end{center} 
\end{figure}

In the case of the stability analysis around thermal 
equilibrium,
the instability observed in the Vlasov equation 
coincides with those of
the mechanical instability and is thus related to the 
occurrence of a first
order phase transition.
Fig.~\ref{rate_fig} shows the contours of the growth 
rate $\Gamma/k$ in
the $T$-$\rho_B$ plane for the phenomenological parton 
model. 
The region of $\Gamma /k >0$
corresponds to the mechanical instability region studied 
in the previous
section.
The analysis based on the Vlasov equation thus 
reproduces the results
of the thermodynamic analysis but goes further in that 
it also gives
values of the growth rates in the unstable region.
 In the instability region close to the phase
boundary, where we trust the model, calculated growth 
rates are 
of the order  $\Gamma/k=0.1$.
For $k=1.0$~fm$^{-1}$ the instability growth time
$1/\Gamma$ is of order  $10$~fm$/c$ and rather small on 
the time scale of
nuclear reactions.

\section{A general result for the case of zero baryon 
density} 
\label{sec_ft}
In this subsection we analyze the situation of finite 
$T$ and
 $\rho_B=0$. We prove that
for any gap equation which yields  $dm/dT<0$, the Vlasov 
equation is stable
against  perturbations of the homogeneous equilibrium 
distribution.
For $T>0$ and $\rho_B=0$ 
the dependence $m(T)$ of the mass on the temperature is 
obtained from the
equilibrium gap equation
\begin{equation}
\frac{dV}{dm}=-\sum_i g_i\int\frac{d^3p}{(2 \pi)^3} 
\frac{m}{E} F_i(E/T) \ ,
\end{equation}
where $F_i(E/T)$ is the equilibrium distributions
for $\mu=0$ written as a function of
 the variable $E/T$ and
\begin{equation}
\label{tuli}
\frac{d^2V}{dm^2}=- \sum_i \frac{g_i}{2 \pi^2} \Bigg[
\int_0^{\infty} \frac{p^4dp}{E^3}
F_i(E/T)  + \int_0^{\infty} \frac{m}{T E}
\bigg(\frac{m}{E}-\frac{dT}{Tdm}\bigg) F_i^{'}(E/T) 
\Bigg] \ .
\end{equation}
Inserting Eq.~(\ref{tuli}) into the  relation 
(\ref{disp_eq})
a dispersion relation for $\omega$ and $k$ is obtained 
which
  depends on the
derivative $dm/dT$~:
\begin{equation}
\label{disp_equ}
\sum_i g_i \int p^2 dp F_i^{'}(E/T) 
\Bigg(\frac{m^2}{E^2} 
\bigg(1-\chi\bigg(\frac{\omega}
{kv}\bigg)\bigg) -\frac{m}{T}\frac{dT}{dm}\Bigg) = 0 \ .
\end{equation}
For any thermal distribution $F(E/T)$ its derivative
$F^{'}(E/T)$ is always nonzero and negative.
The  solutions for the collective
frequency $\omega$
 depend on the sign of 
$dm/dT$. Using the properties of the function 
$\chi(\omega/kv)$ for real
and imaginary $\omega$ \cite{peth} one obtains
\begin{itemize}
\item In the case of $dm/dT > 0$, an imaginary solution 
of the
dispersion cannot be excluded. However, 
 no solution has been found for a version of the 
phenomenological parton model
studied in Ref.~\cite{transp}, where
the thermal mass increases at high temperatures. The 
absence of a solution
can be traced  to the fact
that the thermal mass increases  at high
temperature only like
\begin{equation}
{dm \over dT} < {m \over T}.
%\frac{dm}{dT} < \frac{m}{T}.
%\frac{m}{T}\frac{dT}{dm}>1 \ .
\end{equation}
\item In the case of $dm/dT < 0$, there is no real or 
imaginary
solution of the dispersion relation, and therefore the 
system is stable.
This case applies to the two versions of the 
quasi-particle models studied in this
paper, since $dm/dT<0$ (Figs.~\ref{mtNJL_fig} and 
\ref{mt_fig}).
Any initial perturbation of the
homogeneous distribution will be Landau damped and any
oscillations or instabilities do not propagate or grow.
\end{itemize}

Note that the discussion  is quite general. It applies
 to any  effective quasiparticle theory of the type 
described in 
Sect.~\ref{model_sec},
independently on the statistics of the particles and on 
the specific
dependence of the mass on the temperature. In particular 
we confirm 
 the results of Ref.~\cite{woj} for the 
Nambu-Jona-Lasinio model at
finite temperature and $\rho_B=0$, where no 
instabilities have been found.

\section{Linear instabilities in an  expanding system}

In the previous sections the instabilities of the Vlasov 
equation
are studied at thermal equilibrium within the NJL and 
the
phenomenological parton models. 
While this analysis is important in
order to have a bridge to the condition of mechanical 
instability
in equilibrium thermodynamics, it may not apply to the 
case of
an expanding plasma, since situations far from 
equilibrium may occur, if
thermalization is too slow. 
In the following we neglect thermalization altogether. 
We study 
the instabilities of a collisionless expanding plasma
described by the Vlasov equation. 
The expansion of a
thermal system quickly develops into a non-thermal 
distribution for which
the instabilities may be of different nature. They 
appear also for
$\rho_B=0$. We  study two examples of solutions for the 
Vlasov
equation, the Hubble and Bjorken scenarios.

\subsection{Three-dimensional expansion}

\begin{figure}[b] 
\begin{center}
\epsfxsize=6in
\epsffile{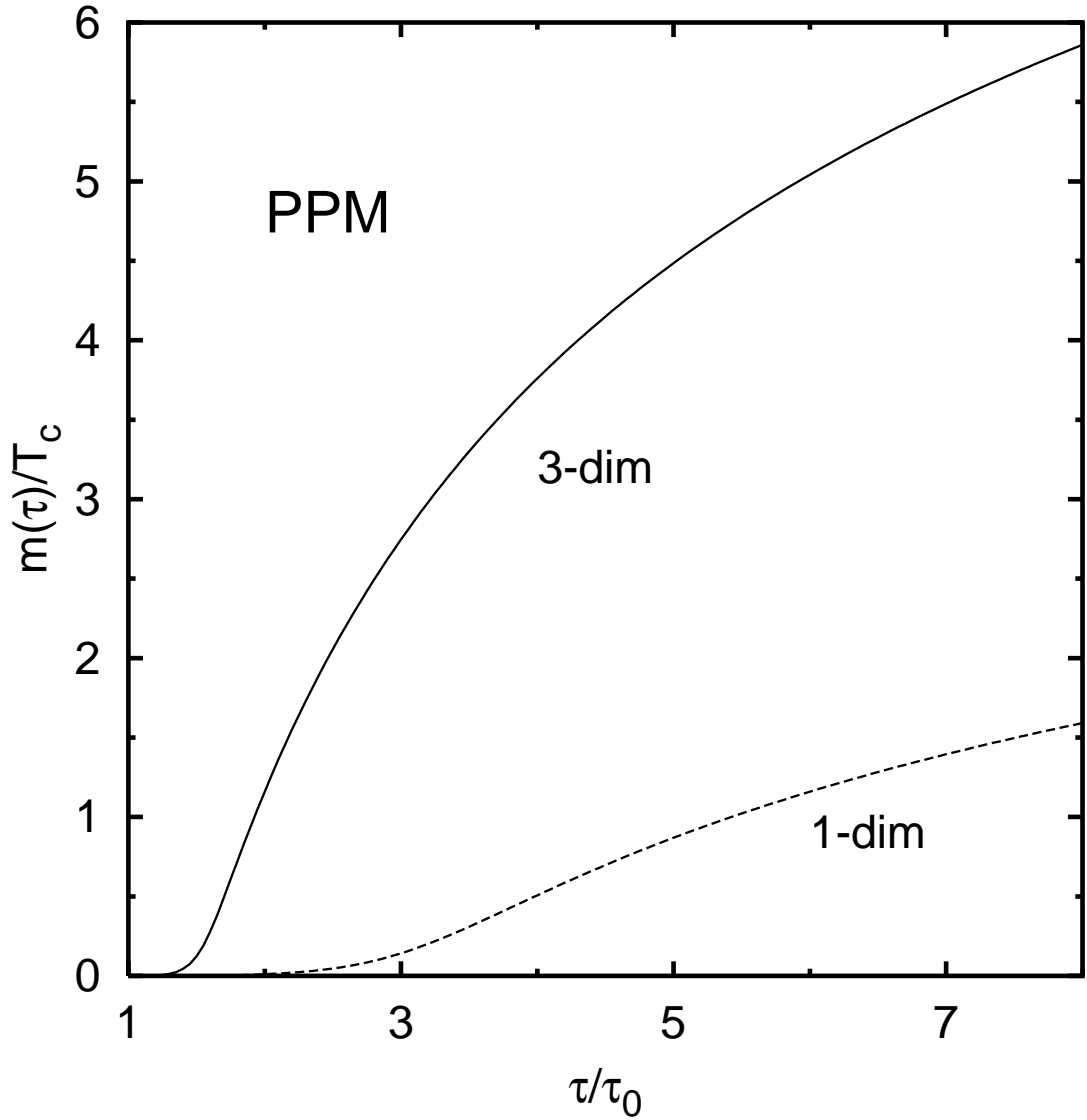}
\caption[]{The parton mass as a function of the proper 
time for the
three-dimensional homogeneous expansion (solid line) and 
for the
Bjorken expansion (dashed line). 
At $\tau / \tau_0 =1$ one has a thermal distribution 
with $T=2T_c$.}
\label{m3d_fig}
\end{center} 
\end{figure}

The solution for an expansion in three dimensions, which 
is
studied in this subsection, is called the Hubble 
solution,
since it describes a homogeneous expansion where the 
local expansion
velocity $u_\mu$ is proportional to the space-time 
vector $x_\mu$. 
For the three-dimensional expansion of the plasma
any function
\begin{equation}
f(p,x,t)=f_0(s) 
\end{equation}
of the variable
\begin{equation}
s=\frac{\tau}{\tau_0}\sqrt{(p^\mu u_\mu)^2-m(\tau)^2} 
\end{equation}
solves the Vlasov equation,
and the mass depends only on the proper time 
$\tau=\sqrt{t^2-{\vec
x}^2}$ and $u_\mu=x_\mu/\tau$ is the local expansion 
velocity \cite{cs}.
If  the system is in thermal
equilibrium at the initial proper time $\tau_0$ with 
mass $m(\tau_0)$
then the solution can be written in the form
\begin{equation}
f(p,x,t)=F(\sqrt{s^2+m(\tau_0)^2}/T)  \ ,
\end{equation}
where $F(x)= {\exp (-x)}$ is the thermal distribution 
function
(In this section we restrict ourselves only to
Boltzmann statistics, since we consider the case 
$\mu=0$. We have
checked that the stability condition and the instability 
growth rates are
not much affected by the statistics.).
For the case $m(\tau_0)=0$, $f$ is only a function of 
$s/T$ for all
space-time points.
In the local rest frame ${\vec x}=0$, $\tau=t$  one has
$s=|{\vec p}|\tau/\tau_0$ and
%for all times. With the initial condition
%$m(\tau_0)=0$ one has
\begin{equation}
\label{zerosol}
f(p,\tau)=F(|{\vec p}|\tau/T\tau_0)
\end{equation}
for all times.
This distribution is a thermal one with time-dependent 
temperature
$T(\tau)=T\tau_0/\tau$, provided the mass remains zero. 
However the
self-consistent mass $m(\tau)$ calculated from the gap 
equation
\begin{equation}
\label{gap3d}
\frac{dV}{dm}=-g\int\frac{d^3p}{(2 \pi)^3} \frac{m}{E}
F(p\tau/T\tau_0) \ 
\end{equation}
increases rapidly with time $\tau$ (Fig.~\ref{m3d_fig}).
 In this case $F(p\tau/T\tau_0)$ is no more
a thermal distribution.
In order to analyze the stability of the solution
(\ref{zerosol}) 
we linearize the Vlasov  and the gap 
(\ref{gap3d}) equations.
 The procedure leads to a complicated expression, since 
both the
 zero order solution $F(s/T)$ and the linear 
perturbation $\delta f$
depend on space and time.
If we restrict ourselves to the vicinity of the point 
$\vec{x}=0$, the
linearized Vlasov equation simplifies, since  the 
$\vec{x}$-dependence 
of the zero order solution is weak:
\begin{equation}
{\partial_t} \delta f(p,x,t)+
\frac{\vec p}{E(\tau)}\cdot {\nabla_x}\delta f(p,x,t) -
\frac{m(\tau)}{E(\tau)}
\nabla_x \delta m(x,t)\cdot \nabla_pF(p\tau/T\tau_0)
=0 \ ,
\end{equation}
where now $\tau=t$.
If one wants to proceed as in Sect.~\ref{linstab} and 
introduces the
Fourier transforms of $\delta f$ and $\delta m$ with 
respect to $\vec x$
and $t$, one has to neglect the time dependence of the
zero order solution $F(p\tau/T\tau_0)$ and $m(\tau)$  in 
comparison to
the expected stronger time variation of the perturbation 
$\delta
f(p,x,t)$. This is equivalent to an adiabatic 
approximation considering
$\tau$ as an external parameter which changes the 
properties of the medium.
After Fourier transformation one obtains the familiar 
form of the
linearized Vlasov equation
\begin{equation}
(\omega-{\vec k}\cdot {\vec v})\delta f(p,k,\omega) + 
\frac{m}{p} {\vec
k}\cdot {\vec v} \partial_p
F(p\tau/T\tau_0) \delta m =0 \ .
\end{equation}
Together with the gap equation (\ref{gap3d}) one gets 
the dispersion
relation in the form:
\begin{equation}
\label{nonth_disp}
\int_0^{\infty}  \frac{p}{E} dp F^{'}(p\tau/T\tau_0) 
\Bigg( 
\chi\bigg(\frac{\omega}{kv}\bigg) 
-p^2\frac{d\ln(\tau/\tau_{0})}{mdm}\Bigg) = 0 \ ,
\end{equation}
which 
 has an imaginary solution for
the frequency if
\begin{equation}
\int_0^{\infty}  \frac{p}{E} dp F^{'}(p\tau/T\tau_0) 
\Bigg( 
1 -p^2\frac{d\ln(\tau/\tau_0)}{mdm}\Bigg) > 0 \ .
\end{equation}

\begin{figure}[b] 
\begin{center}
\epsfxsize=6in
\epsffile{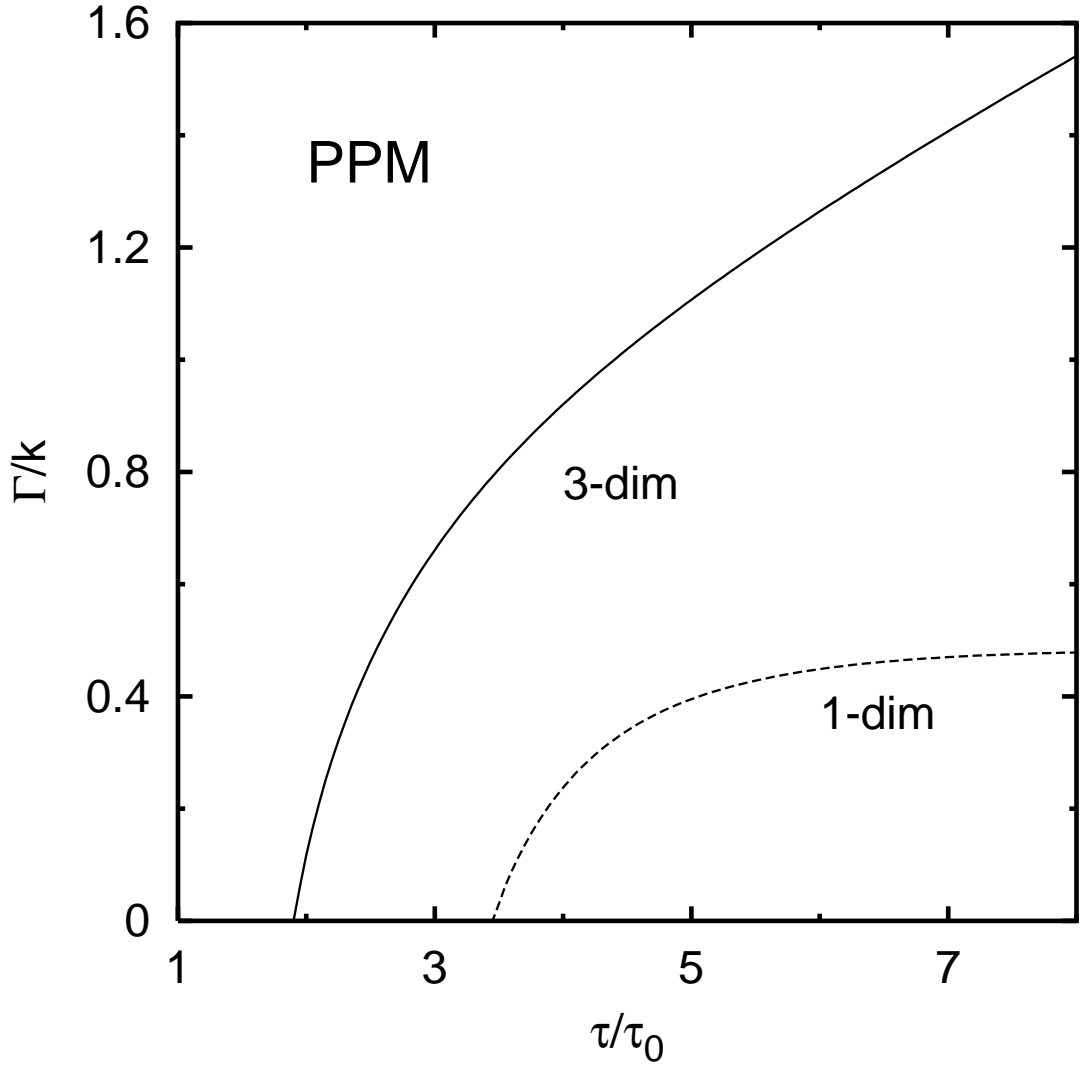}
\caption[]{The instability growth rate $\Gamma/k$ for a 
perturbation of the
three-dimensional homogeneous expansion (solid line) and 
for a longitudinal
perturbation in the Bjorken expansion (dashed line)
  as a function of the proper time. Initial conditions 
like in 
Fig.~\ref{m3d_fig}.}
\label{nonth_fig}
\end{center} 
\end{figure}

The dispersion relation 
 (\ref{nonth_disp}) determines the ratio $\omega/k$. For 
$\Gamma_k>0$
one has an exponentially growing instability.
 Fig.~\ref{nonth_fig} shows the growth rate of the 
instability as a function of
 the expansion time $\tau/\tau_0$.
If we chose a typical value $k=1.0$~fm$^{-1}$, the 
growth rate refers to a
spinodal  instability of the size $1$~fm. At 
$\tau/\tau_0 =1.9$ the system
 becomes unstable and quickly reaches a regime of rather 
short instability
 time scales of order $1/\Gamma= 1.0$~fm$/c$. 

A necessary condition for the appearance of the linear 
instabilities in
the expanding plasma is that the mass increases 
significantly around
$T_c$. For example we
have found that the linear instabilities do not appear 
for the NJL model or for the
three-dimensional Hubble
expansion in the phenomenological parton 
model where the parton mass in 
vacuum is taken smaller than $\sim 3T_c$.
Note that an instability develops for the 
case $\rho_B=0$, although there is no 
  first order phase transition as discussed above.

\subsection{One-dimensional expansion}
For a  one dimensional homogeneous  expansion of the 
plasma 
(Bjorken scenario)
the
Vlasov equation is solved by any function of the scaling 
variable $s$
and of the transverse momentum 
$p_\bot$ \cite{cs}
\begin{equation}
\label{z1d}
 f(p,x,t)=f_0(s,p_\bot) \ ,
\end{equation}
with 
\begin{equation}
s=\frac{\tau}{\tau_0}\sqrt{(p^\mu 
u_\mu)^2-p_\bot^2-m(\tau)^2} \ 
\end{equation}
 and $u_\mu=(t,0,0,z)/\tau$.
In the local rest frame ($z=0$) and for a thermal 
initial state with
$m(\tau_0)=0$ the one-dimensional solution takes the 
form
\begin{equation}
f(p,x,t)=F(\sqrt{(\tau p_\|/\tau_0)^2 
+p_\perp^2}/T)=F(E_\tau/T) \ .
\end{equation}
The mass $m(\tau)$ can be calculated from the gap 
equation and is shown in
Fig.~\ref{m3d_fig}.
In the one-dimensional case  the mass increases more 
slowly than in the
three-dimensional case
since only the longitudinal momentum is dilated.

Linearizing around the one-dimensional homogeneous 
expansion,
we consider two kinds of perturbation, longitudinal 
$\delta f(p,z,t)$
and transverse $\delta f(p,x_\perp,t)$ ones. Any general 
linear solution
of the Vlasov equation  can be factorized into the 
transverse and the
longitudinal ones.
The approximations in deriving the dispersion relation 
from the
linearized
Vlasov and gap equations are the same as in the 
three-dimensional case.
One obtains
\begin{equation}
\int_0^\infty p_\perp dp_\perp \int_{-\infty}^\infty 
dp_\|
\frac{F^{'}(E_\tau/T)}{E E_\tau}\Bigg[\frac{k_\| 
v_\|}{\omega -k_\| v_\|}
+p_\|^2 \frac{d \ln{\tau/\tau_0}}{m dm} \Bigg] = 0 \ ,
\end{equation} 
where $v_\|=p_\|/E$ and $k_\|$ is the longitudinal 
wave-vector of the
perturbation.
The above equation has a purely imaginary solution 
$\omega=\pm i\Gamma$ if
\begin{equation}
\label{stlo}
\int_0^\infty p_\bot dp_\bot \int_{-\infty}^\infty dp_\|
\frac{F^{'}(E_\tau/T)}{E E_\tau}\Bigg[1
-p_\|^2 \frac{d \ln{\tau/\tau_0}}{m dm} \Bigg] > 0 \ .
\end{equation} 
In the transverse direction the dispersion relation 
reads
\begin{equation}
\int_0^\infty p_\perp dp_\perp \int_0^{2\pi} 
d\phi\int_{-\infty}^\infty dp_\|
\frac{F^{'}(E_\tau/T)}{E E_\tau}\Bigg[\frac{k_\perp 
v_\perp \cos{\phi}}
{\omega -k_\perp v_\perp \cos{\phi}}
+p_\|^2 \frac{d (\tau/\tau_0)^2}{ dm^2} \Bigg] = 0 \ ,
\end{equation} 
where $k_\perp$ is the transverse wave-vector of the 
perturbation.
The transverse dispersion relation has a purely 
imaginary frequency
solution if
\begin{equation}
\label{sttr}
\int_0^\infty p_\perp dp_\perp  \int_{-\infty}^\infty 
dp_\|
\frac{F^{'}(E_\tau/T)}{E E_\tau}\Bigg[1 -
p_\|^2 \frac{d (\tau/\tau_0)^2}{ dm^2}  \Bigg] > 0 \ .
\end{equation} 
Comparing the conditions for the longitudinal and
transverse instabilities (Eqs.~(\ref{stlo}) and 
(\ref{sttr}))
 one finds that the system becomes unstable first
in the longitudinal direction, since 
\begin{equation}
\frac{d (\tau/\tau_0)^2}{ dm^2}>\frac{d 
\ln{\tau/\tau_0}}{m dm} \ .
\end{equation}
Furthermore the longitudinal instability growth rate is 
larger than the
transverse one.
 Fig.~\ref{nonth_fig}  shows the time dependence of the
longitudinal instability growth rate for the 
one-dimensional expansion.
The growth rate $\Gamma$
 is much smaller than in the three-dimensional case and 
is
comparable to the growth rates in equilibrium at finite 
baryon density.
For the wave-vector $k=1.0$~fm$/c$
one has a maximum value of $\Gamma \simeq 0.5~c/$fm at 
$\tau/\tau_0=8$.
The transverse perturbation remains stable for the 
phenomenological
parton model.

\section{Clusterization in a molecular dynamics 
calculation}

\label{md_sec}

When a perturbation of the zero order Vlasov equations 
is 
unstable, the linear approximation breaks down for 
larger times.
The development of the instabilities beyond the linear
approximation can only be studied numerically.
In the case of the mean field evolution 
(Vlasov equation), 
the numerical noise due to the finite number of test 
particles
is a big problem
\cite{bl}. It is very difficult to disentangle the 
numerical noise from
the true fluctuations of the system acting as a seed for 
the 
development of the unstable modes.
We have not attempted to solve this problem. Instead we 
study
a different but closely related
model. It is the molecular dynamical model with medium 
dependent
masses. The local mean-field approximation to this model 
reduces to the
Vlasov equation
studied in the previous section. The model deals with  
classical partons whose masses are related to the 
density in the
vicinity. The time dependent positions $\vec{x}_i(t)$ 
and momenta $\vec{p}_i(t)$ 
of the $N$ partons follow Hamilton's equations
\begin{eqnarray}
\frac{d\vec{p}_i}{dt}  &=&-\frac{m_i}{E_i}\nabla_x m_i 
\nonumber \\
\frac{d\vec{x}_i}{dt} & = & \frac{\vec{p}_i}{E_i}  \ .
\end{eqnarray}
The parton mass is calculated from the gap equation with
the scalar density in its vicinity. 
This requires the introduction of a range over which the 
density is
calculated. 
We define the scalar density at position 
${\vec{x}_i}(t)$
at time $t$ by
\begin{equation}
\rho_s(\vec{x}_i(t))={1 \over (2 \pi \sigma)^{3/2}}
  \sum_{j \ne i} 
\frac{m_j}{E_j}e^{-(\vec{x}_i-\vec{x}_j)^2/2 \sigma^2} \ 
,
\end{equation}
where the sum runs over all the particles $j$ (with 
positions $\vec{x}_j(t)$,
masses $m_j(t)$ and
energies
$E_j(t)$) other than the parton $i$.
It determines the mass of the parton $i$ through the gap 
equation
of the phenomenological parton model
\begin{equation}
\frac{dV(m_i)}{dm}=-\rho_s(\vec{x}_i(t)) \ .
\end{equation}
The introduction of the finite range $\sigma$ (which we 
take 
$\sigma=0.5$~fm) has important consequences:
(i) For a finite value of $\sigma$, the mass of particle 
$i$
at $\vec{x}_i$ and $t$ cannot be determined by the 
masses of the particles $j$
at the same time $t$, but only at earlier times 
(causality). As long as 
the masses vary slowly with time, the violation of 
causality may have no 
dramatic consequences. (ii) The size of the clusters 
into which the system 
breaks up will be strongly influenced by the value of 
$\sigma$,
but we have not yet studied this dependence. 

\begin{figure}[b] 
\begin{center}
\epsfxsize=6in
\epsffile{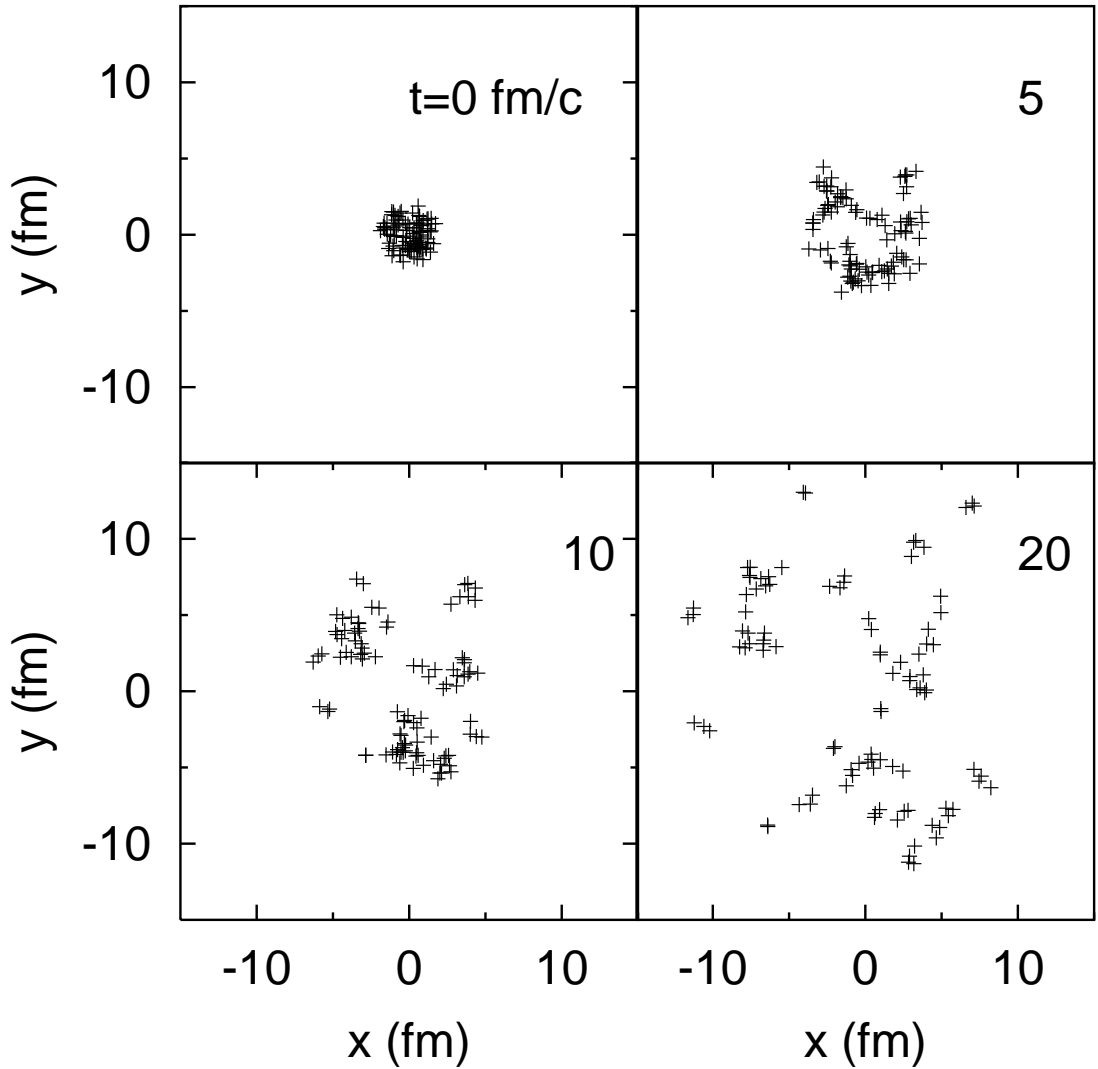}
\caption[]{Expansion of a fireball of partons calculated 
within 
molecular dynamics. The positions of the partons with 
coordinates $x$ and $y$ are projected on the plane 
$z=0$. Each cross
represents a parton.
At $t=20$~fm/c there is no cluster with a single 
parton.}
\label{md_fig}
\end{center} 
\end{figure}

Using the gap equation for the phenomenological parton 
model,
we have performed a 
simulation using 100 partons with initial temperature 
$T=180$~MeV 
distributed in a fireball of radius $r_0=2$~fm. The 
system 
expands and its density drops. 
Fig.~\ref{md_fig} presents the evolution of the system 
at different times
by showing the positions ($x$, $y$) of the
partons projected on the $z=0$ plane.
The partons tend to stay grouped in regions and 
structures of large
local density. Eventually as the expansion proceeds the 
partons
group themselves  into compact clusters composed of at 
least two
partons. 
There are no single partons. This is as close as one can 
do
with the hadronization process in a model which does not 
know
of color and not of quantum mechanics.
The qualitative picture obtained from this calculation 
corresponds to
the final stage of the dynamical break up of the system, 
which manifest
itself first as a linear instability studied in the 
previous
sections. The molecular dynamics evolution follows the 
system 
for times beyond the applicability of the linear 
response.
%It shows that the system forms compact clusters which then can separate
%from one another, allowing for a substantial collective flow.
%These clusters could then decay into physical hadrons in the second
%stage of the hadronization.

\section{Summary and conclusions}
We have studied the instabilities of a plasma of 
partons. The plasma is 
treated in a quasi-particle model with an effective mass 
$m$, which 
depends on $T$ and $\mu$ in the equilibrium situation 
and on
$\vec{x}$, $t$ for the expansion. The gap equation which 
relates the mass
to the scalar density of the partons defines the 
underlying dynamics. We
have studied two cases in parallel:  (i) a gap equation 
derived from the 
NJL-Lagrangian in the mean field approximation and (ii) 
a phenomenological
gap equation which is chosen so that certain observables 
calculated from
the lattice are reproduced by the quasi-particle model. 
While both
models show a crossover transition for $\mu=0$ and a 
first order phase 
transition for $\mu\neq 0$, only the phenomenological 
model has 
properties of confinement, in that the parton mass 
becomes very large for 
vanishing density. The study of these two, in some 
sense, complementary 
models, permits to get an idea on the model dependence 
of the results. 
The stability analysis has been performed on four 
levels:

\begin{itemize}
\item For thermal equilibrium, one calculates the 
regions in the 
$T-\rho_B$ plane, in which the system is stable, 
metastable and unstable.
These regions are calculated starting from the pressure 
as 
a function of $T$ and $\rho_B$. At thermal equilibrium 
instabilities are
expected only at and due to the first order phase 
transitions.
Since both models, the NJL and the phenomenological one
show a first order phase transition for certain values 
of $T$ and
$\rho_B$, the phase diagrams look rather similar from a 
qualitative 
point of view, but the values of $T$ and $\rho_B$ at the 
phase
boundaries are rather different.

\item Another interesting result found from the analysis of 
the phase diagram of the phenomenological parton model 
is the presence of the first order phase transition
at finite baryon densities. Although we have followed 
the lattice QCD data at zero baryon density which show 
only a crossover transition arround $T_c$ the 
generalization to finite baryon density using a gap 
equation depending only on the scalar density leads to a 
first order phase transition. This simple model shows 
that the occurence of a first order phase transition  
in the highly excited matter with finite baryon density 
created in ultrarelativistic heavy ion collisions
cannot be excluded from the lattice QCD data at zero 
baryon density.

\item The growth rates for the instabilities found in 
the
thermodynamic analysis can be  calculated from the 
Vlasov equation in 
a linear response analysis around thermal 
equilibrium. The neglect of the collisions in the linear 
response analysis in this work is justified when the 
instability 
growth rate is larger than the equilibration rate.
We prove that for any quasi-particle model for which 
$dm/dT<0$, one has a stable system for $\rho_B=0$. 
In general
the dispersion relation for frequency $\omega$ and wave 
vector $k$
of the instability has no scale and gives $\omega 
\propto k$,
since the gap equation corresponds to a zero range 
interaction.
In the region of the mixed phase obtained from 
equilibrium thermodynamics,
one finds solutions with Im$\omega>0$, {\it i.e.}, 
perturbations which grow 
exponentially in time. For the typical scale 
corresponding to the
 wave vectors $k\simeq 1.0$~fm$^{-1}$ and as long as one 
is not too far
away from the phase boundary\footnote{The models based 
solely on the parton
 degrees of freedom may have not much physical
meaning far in the confined phase.}
the value of the growth rate Im$\omega \simeq 
0.1$~(fm$/c$)$^{-1}$ is surprisingly
small (compared to the expansion time scale of the 
plasma).

\item The stability behavior far from equilibrium has 
been studied by
investigating the Hubble- and Bjorken-solutions, 
({\it i.e.}, three- and one-dimensional expansion, 
respectively)
of the Vlasov equation again by the method of linear 
response. 
The condition for the validity of this procedure is that 
the growth
rate of the instability is faster than the expansion 
time
(which itself must be larger than the relaxation time in 
order 
to be able to use the collisionless
Vlasov equation).
No instabilities are found for the NJL-model, 
while the phenomenological model quickly ($\tau \geq 
2\tau_0$) 
develops instabilities with significant growth
rates (Im$\omega \simeq 0.5$ to 1.0~(fm$/c)^{-1}$ for 
$k=1.0$~fm$^{-1}$)
even for $\rho_B=0$. We call these instabilities 
dynamical ones,
since they are obviously not related to the first order 
phase 
transition but  rather to the confinement properties of 
the phenomenological
quasi-particle model.

\item The analysis of the instabilities based on linear 
response, be it at thermal equilibrium or for an 
expansion far
from equilibrium, is limited to small times, if one 
finds exponential
growth. The behavior over large times can only be 
studied 
numerically.
We have attempted a first step in this direction, in 
that we 
have followed the expansion of a plasma described by the 
phenomenological 
quasi-particle model in a molecular dynamical 
calculation.
We find indeed that the expansion of the system 
eventually leads to clusters
with at least two partons. This is as far as one can get 
in a classical
model which does not know color degrees of freedom.
More work in this direction is intended.

\end{itemize}

\acknowledgments
The authors thank Hilmar Forkel and G\'{a}bor Papp for 
several helpful 
discussions. P.B. thanks the A.v. Humboldt foundation 
for financial
support. Y.B.H. is supported by the grant 06~HD742 from 
the 
German Federal Ministry of Education and Research.
J.H. is grateful to A. Gal and the Physics Department of 
the 
Hebrew University for their hospitality and support 
during a stay where
some this work has been done.

%\begin{thebibliography}{99}

%\end{thebibliography}

\newpage

\end{document}